\documentclass[12pt]{article}

\usepackage{sbc-template}
\usepackage{graphicx,url}
\usepackage[utf8]{inputenc}
\usepackage[english]{babel}
\usepackage{multirow}
\usepackage{graphicx}
\usepackage{float}
\usepackage{multirow}
\usepackage[nolist]{acronym}
\usepackage{amsmath}
\usepackage{subcaption}
\usepackage{tikz}
\usepackage{colortbl}
\usepackage{xcolor}
\usepackage{booktabs}

\title{Optimizing Edge Gaming Slices through an Enhanced User Plane Function and Analytics in Beyond-5G Networks}

\author{Bruno Marques da Silva\inst{1},  Larissa Ferreira {Rodrigues Moreira}\inst{1}, \\Flávio {de Oliveira Silva}\inst{2} and Rodrigo Moreira\inst{1}}

\address{Institute of Exact and Technological Sciences -- Federal University of Viçosa
  (UFV)\\
  Rio Paranaíba -- MG -- Brazil
\nextinstitute
  Department of Informatics -- School of Engineering\\
  University of Minho (UMinho) -- Braga -- Portugal
  \email{\{bruno.silva63, rodrigo, larissa.f.rodrigues\}@ufv.br, flavio@di.uminho.pt}  
}

\begin{document} 
\acrodef{3GPP}{3rd Generation Partnership Project}
\acrodef{AI}{Artificial Intelligence}
\acrodef{AMF}{Access and Mobility Management Function}
\acrodef{ANOVA}{Analysis of Variance}
\acrodef{API}{Application Programming Interface}
\acrodef{AUSF}{Authentication Server Function}
\acrodef{AUC}{Area Under the Curve}
\acrodef{B5G}{Beyond Fifth Generation}

\acrodef{CPU}{Central Processing Unit}
\acrodef{CHF}{Core Charging Function}
\acrodef{DoS}{Denial of Service}
\acrodef{DDoS}{Distributed Denial of Service}
\acrodef{DNN}{Deep Neural Network}
\acrodef{DRL}{Deep Reinforcement Learning}
\acrodef{DT}{Decision Tree}
\acrodef{ETSI}{European Telecommunications Standards Institute}
\acrodef{E2E}{End-to-End}
\acrodef{FIBRE}{Future Internet Brazilian Environment for Experimentation}
\acrodef{FIBRE-NG}{Future Internet Brazilian Environment for Experimentation New Generation}
\acrodef{5G}{Fifth-generation of Mobile Telecommunications Technology}
\acrodef{GNN}{Graph Neural Networks}
\acrodef{GTP}{General Packet Radio Service Tunnelling Protocol}
\acrodef{GPU}{Graphics Processing Unit}
\acrodef{HTM}{Hierarchical Temporal Memory}

\acrodef{IAM}{Identity And Access Management}
\acrodef{IID}{Informally, Identically Distributed}
\acrodef{IoE}{Internet of Everything}
\acrodef{IoT}{Internet of Things}
\acrodef{KNN}{K-Nearest Neighbors}
\acrodef{KPI}{Key Performance Indicator}
\acrodef{KPIs}{Key Performance Indicators}
\acrodef{LSTM}{Long Short-Term Memory}
\acrodef{LMM}{Linear Mixed Model}
\acrodef{LOL}{League of Legends}
\acrodef{M2M}{Machine to Machine}
\acrodef{MAE}{Mean Absolute Error}
\acrodef{ML}{Machine Learning}
\acrodef{MOS}{Mean Opinion Score}
\acrodef{MAPE}{Mean Absolute Percentage Error}
\acrodef{MSE}{Mean Squared Error}
\acrodef{mMTC}{Massive Machine Type Communications}
\acrodef{MFA}{Multi-factor Authentication}
\acrodef{MQTT}{Message Queuing Telemetry Transport}
\acrodef{MEC}{Multi-access Edge Computing}
\acrodef{NF}{Network Function}
\acrodef{NFs}{Network Functions}
\acrodef{NWDAF}{Network Data Analytics Function}
\acrodef{NSSF}{Network Slice Selection Function}
\acrodef{NRF}{Network Repository Function}
\acrodef{NEF}{Network Exposure Function}
\acrodef{OSM}{Open Source MANO}
\acrodef{PDU}{Protocol Data Unit}
\acrodef{PCF}{Policy Control Function}
\acrodef{QoE}{Quality of experience}
\acrodef{QoS}{Quality of Service}
\acrodef{RAM}{Random-Access Memory}
\acrodef{RF}{Random Forest}
\acrodef{RL}{Reinforcement Learning}
\acrodef{RMSE}{Root Mean Square Error}
\acrodef{RNN}{Recurrent Neural Network}
\acrodef{RAN}{Radio Access Network}
\acrodef{RTT}{Round-trip time}
\acrodef{ROC}{Receiver Operating Characteristic}
\acrodef{SDN}{Software-Defined Networking}
\acrodef{SFI2}{Slicing Future Internet Infrastructures}
\acrodef{SLA}{Service-Level Agreement}
\acrodef{SON}{Self-Organizing Network}
\acrodef{SMF}{Session Management Function}
\acrodef{SMSF}{Short Message Service Function}
\acrodef{TEID}{Tunnel Endpoint Identifier}
\acrodef{TFT}{Teamfight Tactics}
\acrodef{UE}{User Equipment}
\acrodef{UEs}{User Equipments}
\acrodef{UDM}{Unified Data Management}
\acrodef{UDR}{Unified Data Repository}
\acrodef{UPF}{User Plane Function}
\acrodef{VoD}{Video on Demand}
\acrodef{VR}{Virtual Reality}
\acrodef{V2X}{Vehicle-to-Everything}
\acrodef{VAL}{Valorant}


\maketitle

\begin{abstract}

The latest generation of games and pervasive communication technologies poses challenges in service management and Service-Level Agreement compliance for mobile users. State-of-the-art edge-gaming techniques enhance throughput, reduce latency, and leverage cloud computing. However, further development of core functions such as the User Plane Function (UPF) is needed for non-intrusive user latency measurement. This paper proposes a closed-loop architecture integrating the Network Data Analytics Function (NWDAF) and UPF to estimate user latency and enhance the 5G control plane by making it latency-aware. The results show that embedding an artificial intelligence model within NWDAF enables game classification and opens new avenues for mobile edge gaming research.
\end{abstract}
     
\section{Introduction}\label{sec:introduction}


Advancements in \ac{5G} networks and the unprecedented capacity of \ac{GPU} have opened up support for bold network metrics to meet the demands of business verticals such as Virtual Reality (VR), Augmented Reality (AR), and entertainment applications such as online gaming~\cite{Shankar2024}. Cloud gaming, where a server streams games to users or devices, presents challenges due to network dynamism, requiring high throughput for streaming \ac{KPIs} and low latency for satisfactory \ac{QoE}~~\cite{Soares2024}.


\ac{AI} plays a crucial role in network slicing management and orchestration, enabling precise service customization for resource-intensive applications~\cite{Moreira2023, RodriguesMoreira2024}. Enhancing Edge gaming in \ac{5G} remains challenging due to the need for coordinated interventions across \ac{RAN}, \ac{UPF}, and cloud services~\cite{Soares2024}. In this context, \ac{AI} supports \ac{QoS} assurance and \ac{SLA} compliance in dynamic environments~\cite{Kougioumtzidis2024}, allowing mobile network control plane mechanisms to enforce strict performance metrics for deployed network slices, particularly in online gaming.


State-of-the-art edge gaming approaches, including \ac{MEC}-based methods, wireless network enhancements, and traffic engineering for \ac{QoS} support~\cite{Soares2024, Shankar2024, Kougioumtzidis2024}, do not incorporate the \ac{NWDAF} function from \ac{3GPP} Release 16, which provides analytics in the \ac{5G} core network. This paper introduces \ac{UPF} instrumentation with a user-space filter to measure \ac{UE} latency and report it to \ac{NWDAF} for core network analysis and potential \ac{SLA} improvements. The main contributions are i) \ac{UPF} instrumentation with user-space filters for slice latency estimation based on \ac{TEID}, ii) empirical evaluation using a real dataset, and iii) performance assessment of \ac{AI} techniques in this domain.


The remainder of this paper is organized as follows: Section~\ref{sec:related-work} reviews prior work on cloud gaming challenges; Section~\ref{sec:method} outlines the proposed method; Section~\ref{sec:experimental_setup} describes the testbed and technologies used; Section~\ref{sec:results_and_discussion} analyzes the results, insights, and lessons learned; and Section~\ref{sec:concluding_remarks} presents conclusions and future work.

\section{Related Work}\label{sec:related-work}

\cite{Slivar2019} addressed the optimization of resource allocation for multiple cloud gaming users sharing a bottleneck link in \ac{5G} networks by employing \ac{QoE}-aware algorithms based on subjective \ac{QoE} models; their regression analysis utilized \ac{MOS} scores from games such as Serious Sam 3, Hearthstone, and Orcs Must Die! Unchained, derived from controlled laboratory experiments.

\cite{Zhang2019} developed the EdgeGame, a framework leveraging mobile edge computing to address high latency and bandwidth consumption in cloud gaming using a deep reinforcement-learning-based algorithm for adaptive bitrate control. The system optimizes \ac{QoE} under dynamic network conditions. 

\cite{Baena2023} proposed a comprehensive dataset containing quality indicators to evaluate video streaming and cloud gaming services' \ac{E2E} performance over \ac{5G} networks, utilizing a regression approach to estimate \ac{E2E} service metrics based on network parameters.

\cite{Rossi2024} evaluated three objective \ac{QoE} prediction models for mobile cloud gaming, leveraging linear, polynomial, and nonlinear regression to address the impact of \ac{QoS} factors including \ac{RTT} and the models were trained and validated on a publicly available dataset derived from controlled subjective tests.

\cite{Carvalho2024} employed transfer learning to address the challenge of cross-domain \ac{QoE} estimation in cloud gaming services, focusing on adaptation from wired to mobile \ac{5G} networks. Their regression-based model significantly reduced the \ac{MSE} by leveraging a dataset of subjective \ac{QoE} assessments collected under varying network conditions.

\cite{Soares2024} proposed an expanded stacking learning model that integrates datasets from wired and mobile network contexts, focusing on wireless networks (5G). The study employed a regression approach and utilized a merged dataset with 3,323 instances from 88 players, combining features from different gaming environments to effectively predict \ac{QoE}.

In contrast to the previously mentioned works, our proposal advances the state-of-the-art by introducing a latency-aware closed-loop architecture within the 5G control plane. By integrating \ac{NWDAF} and \ac{UPF}, we enable nonintrusive latency measurement and leverage \ac{AI} for real-time game classification and latency forecasting. This innovative approach ensures \ac{SLA} compliance, enhances edge-gaming service management, and provides a robust solution to address the complexities of modern gaming environments.

\section{Proposed Method}\label{sec:method}

\begin{figure}[htbp]
    \centering
    \includegraphics[width=0.9\textwidth]{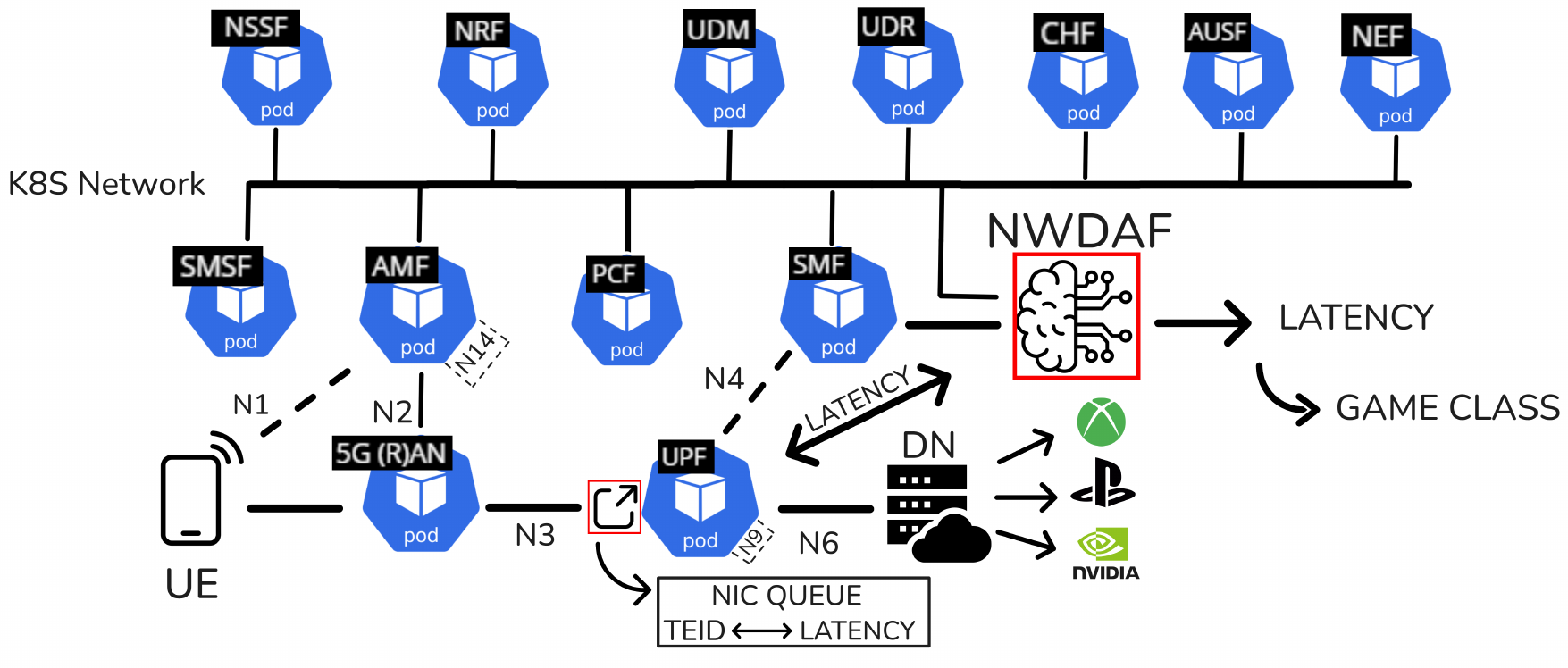}
    \caption{Proposed B5G Architecture for Enhanced Edge Gaming.}
    \label{fig:proposed_method}
\end{figure}

This paper proposes a non-intrusive method for user service evaluation, estimating gaming user experience quality solely through network infrastructure analysis. The method calculates the latency within a specific packet flow window passing through the N3 interface of \ac{UPF}, as illustrated in Figure~\ref{fig:proposed_method}.

The functions of the \ac{5G} core are represented at the top, including \ac{NSSF}, \ac{NRF}, \ac{UDM}, \ac{UDR}, \ac{CHF}, \ac{AUSF}, \ac{SMSF}, \ac{AMF}, \ac{PCF}, \ac{SMF}, and \ac{NEF}, which provide support for session control, policy management, traffic forwarding, and user authentication.


Our method involves instrumenting \ac{UPF} with a user-space latency monitor that measures the temporal offset of packets for each \ac{TEID} in each slice. The latency of each received packet was recorded and submitted to \ac{NWDAF} for analysis. The architecture envisions the \ac{NWDAF} notifying the \ac{SMF} about the quality of service perceived by the user, based on the game class consumed by the \ac{UE} and whether intervention is needed. The framework integrates pretrained \ac{ML} models, including \ac{KNN}, \ac{RF}, \ac{DT}, \ac{LSTM}, and CatBoost.

\begin{figure}[htbp]
    \centering
    \includegraphics[width=0.6\textwidth]{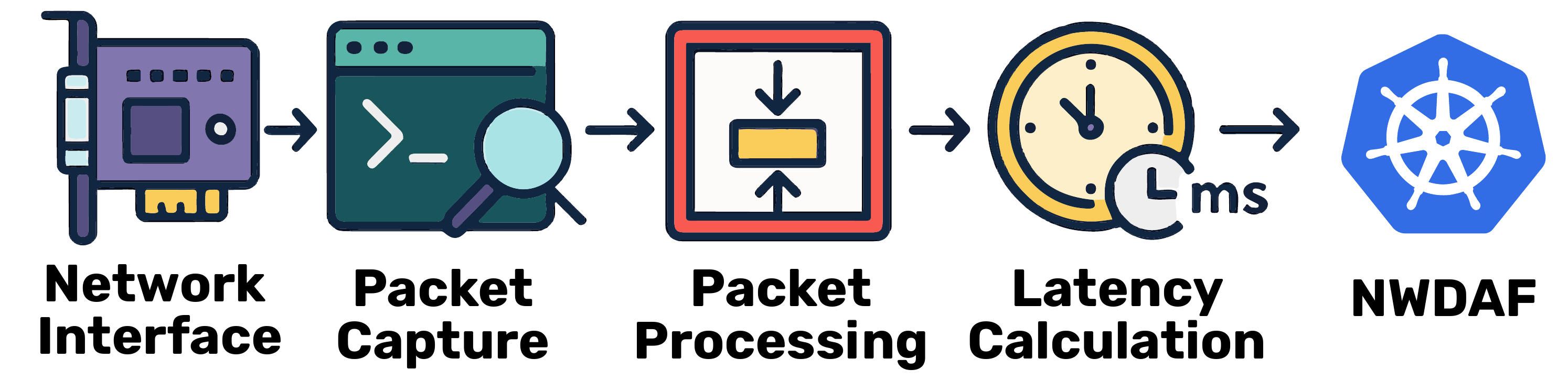}
    \caption{Time-Shift Latency Estimation for Edge Gaming.}
    \label{fig:Time-shift-latency-method}
\end{figure}

Figure~\ref{fig:Time-shift-latency-method} illustrates the latency monitor operating on the N3 interface of \ac{UPF}. The approach analyzes the \ac{UPF} pod interface, capturing packets using specific filters to extract the \ac{TEID}, a 32-bit field in the \ac{GTP} header uniquely identifying the tunnel endpoint. Latency is estimated from the arrival and return timestamps of the packets. Based on this data, \ac{NWDAF} can classify the game by identifying latency patterns.

Time-shift latency estimation measurement involves observing packet timestamps at an \ac{UPF} node without introducing additional traffic. If a packet arrives at the \ac{UPF} at \( t_{\text{in}} \) and departs at \( t_{\text{out}} \), the latency is expressed as:

\[
L = t_{\text{out}} - t_{\text{in}}
\]

For bidirectional communication, the total latency can be calculated as:

\[
L_{\text{total}} = (t_{\text{out}}^{\text{request}} - t_{\text{in}}^{\text{request}}) + (t_{\text{out}}^{\text{response}} - t_{\text{in}}^{\text{response}})
\]

As shown in Figure.~\ref{fig:Time-shift-latency-method}, we construct a filter in the user-space of the \ac{UPF} to associate the estimated latency with the \ac{TEID}, enabling us to estimate the latency for different \ac{UEs} that may be registered in the \ac{5G} core.


Furthermore, our proposed framework assumes an \ac{NWDAF} containing an \ac{API} of pre-trained \ac{ML} models to instruct the \ac{SMF} regarding the state of the \ac{UEs} sessions. Based on the estimated latencies, \ac{NWDAF} identified the type of game played by \ac{UE}. If service degradation is detected, the \ac{SMF} can take corrective actions, such as reconfiguring resource allocation, suggesting routing policy changes, or dynamically adapting network parameters to optimize the user experience.

\section{Experimental Setup}\label{sec:experimental_setup}


This paper instantiates a Fabric testbed virtual machine with 32 GB RAM and 16 vCPUs, running a Kubernetes 1.28 cluster. The free5GC services, including control plane and user plane components, are deployed on this cluster. For the proof of concept, we used a dataset comprising 69,395 instances and 16 features, representing characteristics extracted from various games~\cite{Hassanein2025}. As a classification problem, the target variable corresponds to the game category, consisting of three classes: \ac{LOL}, \ac{TFT}, and \ac{VAL}.


Features encompass numeric and categorical attributes, including source, destination, latitude, and longitude, which capture key aspects of game behavior and performance. These features were structured during preprocessing to enable statistical analysis and model processing for integration into the framework.

\section{Results and Discussion}\label{sec:results_and_discussion}


We evaluated the \ac{UE}-experienced latency by analyzing the time-shifted packet flow through the N3 interface of the \ac{UPF}. Baseline latency, defined as the \ac{UE}-perceived \ac{RTT} when transmitting data, was measured while inducing a sinusoidal synthetic load (1 ms to 600 ms over a 30-second cycle) on the packet recipient. Simultaneously, our filter captured timestamps of packets transiting the N3 interface with a specific \ac{TEID} to estimate intermediate latency.


Table~\ref{tab:measured_estimation_error} shows our non-intrusive method effectively estimates \ac{UE} latencies, with low normalized errors (\ac{MSE}: 0.019, \ac{MAE}: 0.085) and a high $R^2$ (0.980). Despite a higher normalized \ac{MAPE} of 25.090 in some cases, the original \ac{MAPE} of 6.352 highlights the model’s practical adequacy, with an average error of $\approx 6.352$ between actual and estimated latencies via the N3 interface.

\begin{table}[htbp]
\centering
\scriptsize
\caption{\ac{UPF} Latency Estimation Performance.}
\label{tab:measured_estimation_error}
\begin{tabular}{|l|c|}
\hline
\multicolumn{1}{|c|}{\textbf{Metric}}       & \textbf{Value} \\ \hline
\ac{MSE} (Normalized)             & 0.019          \\ \hline
\ac{MAE} (Normalized)            & 0.085          \\ \hline
\ac{MAPE} (Normalized) & 25.090         \\ \hline
R2 Score (Normalized)                       & 0.980          \\ \hline
\ac{MAPE} (Original)   & 6.352          \\ \hline
\end{tabular}
\end{table}

\begin{figure}[htbp]
    \centering
    \begin{tabular}{cc}
			\includegraphics[width=0.54\textwidth]{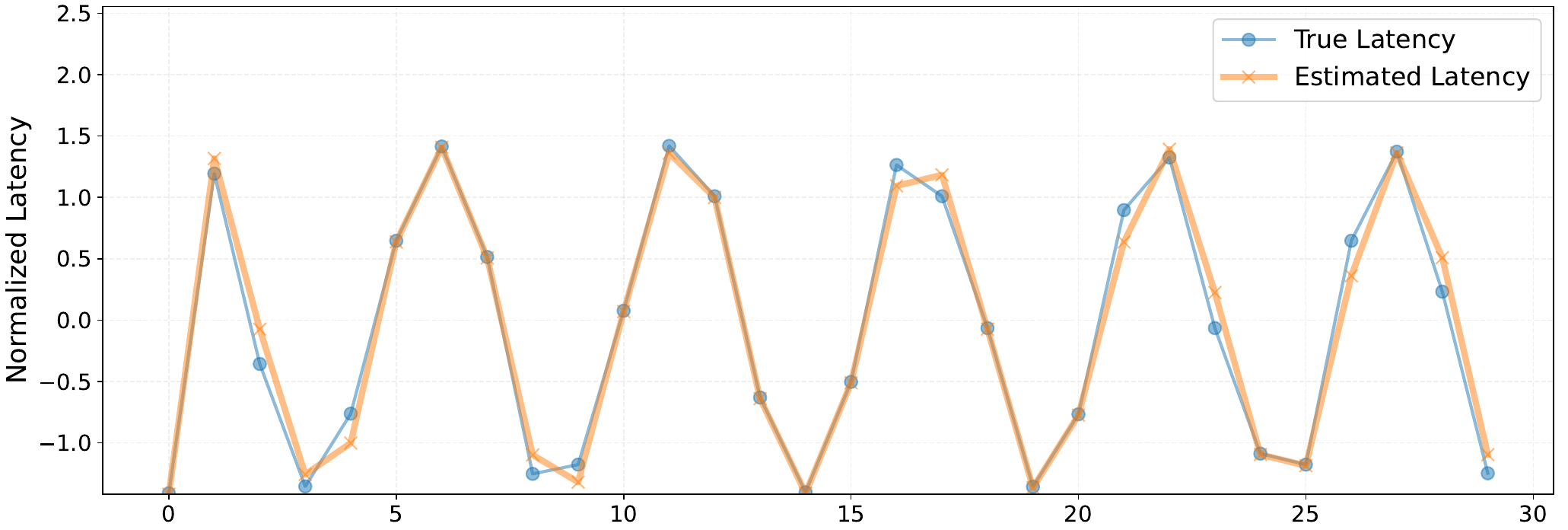} &
			\includegraphics[width=0.4\textwidth]{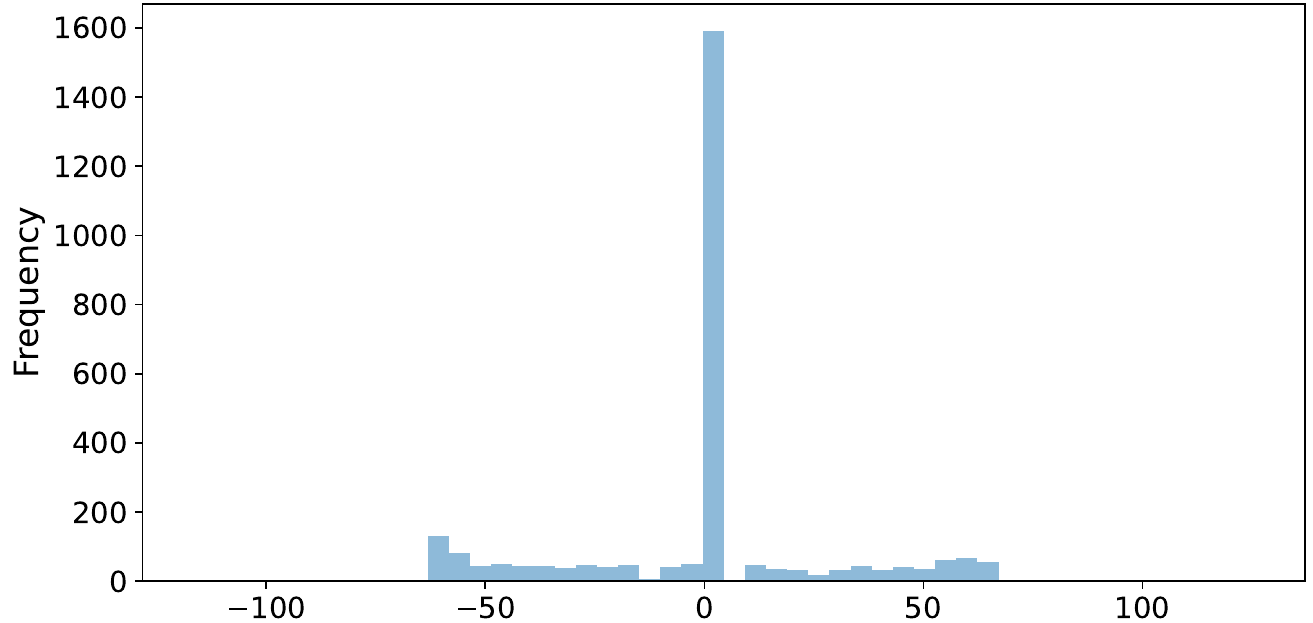} \\
			(a) Sample of Real and Estimated Latency. & (b) Histogram of Errors.\\
    \end{tabular}
    \caption{Estimation Analyze.}
    \label{fig:combined_figures}
\end{figure}

Figure~\ref{fig:combined_figures}a depicts a 10-minute time series sample comparing the latency experienced by the \ac{UE} with the estimated latency in the cluster where the \ac{UPF} was deployed. This reinforces the approach's promise and accuracy, as Figure~\ref{fig:combined_figures}b highlights a significant accumulation of estimation errors near zero.

\begin{figure}[ht]
    \centering
    \begin{tabular}{cc}
        \includegraphics[width=0.40\textwidth]{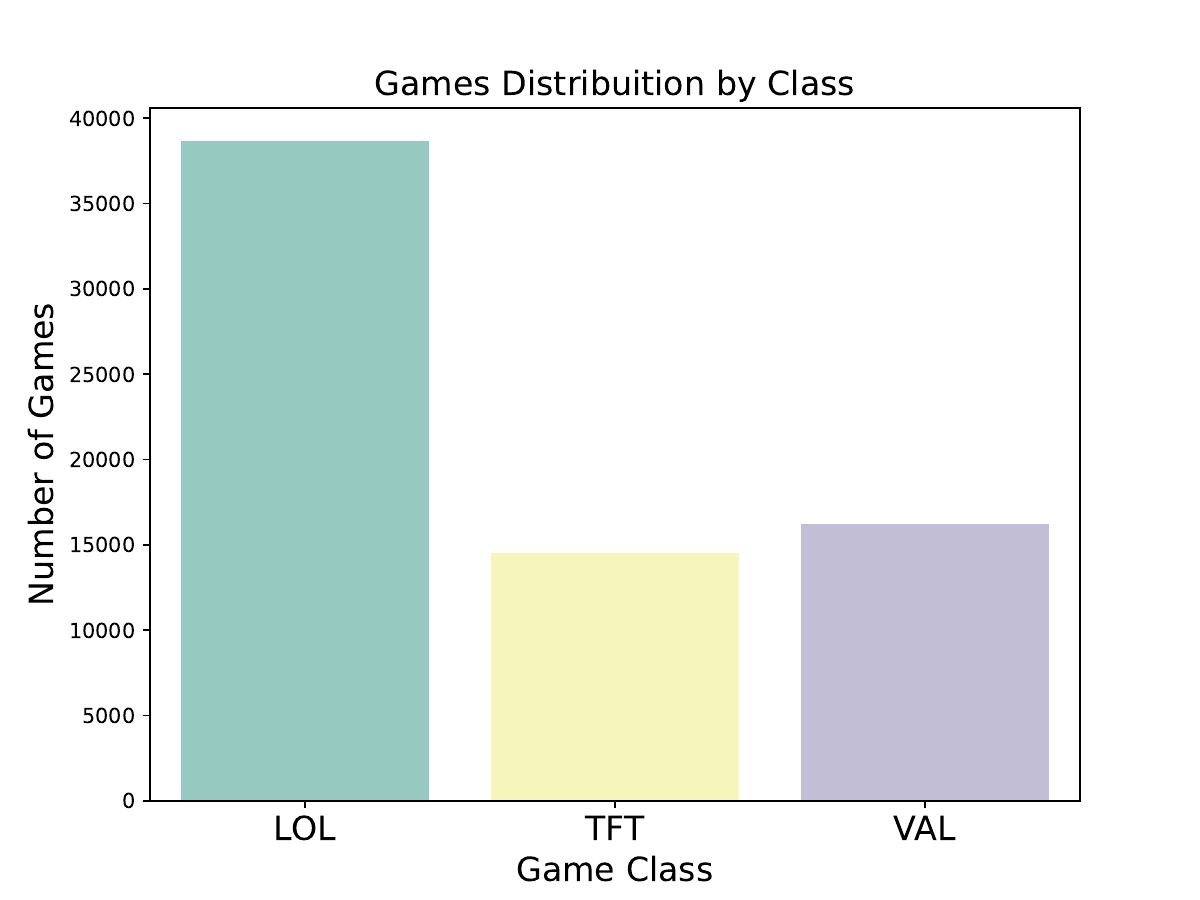} &
        \includegraphics[width=0.40\textwidth]{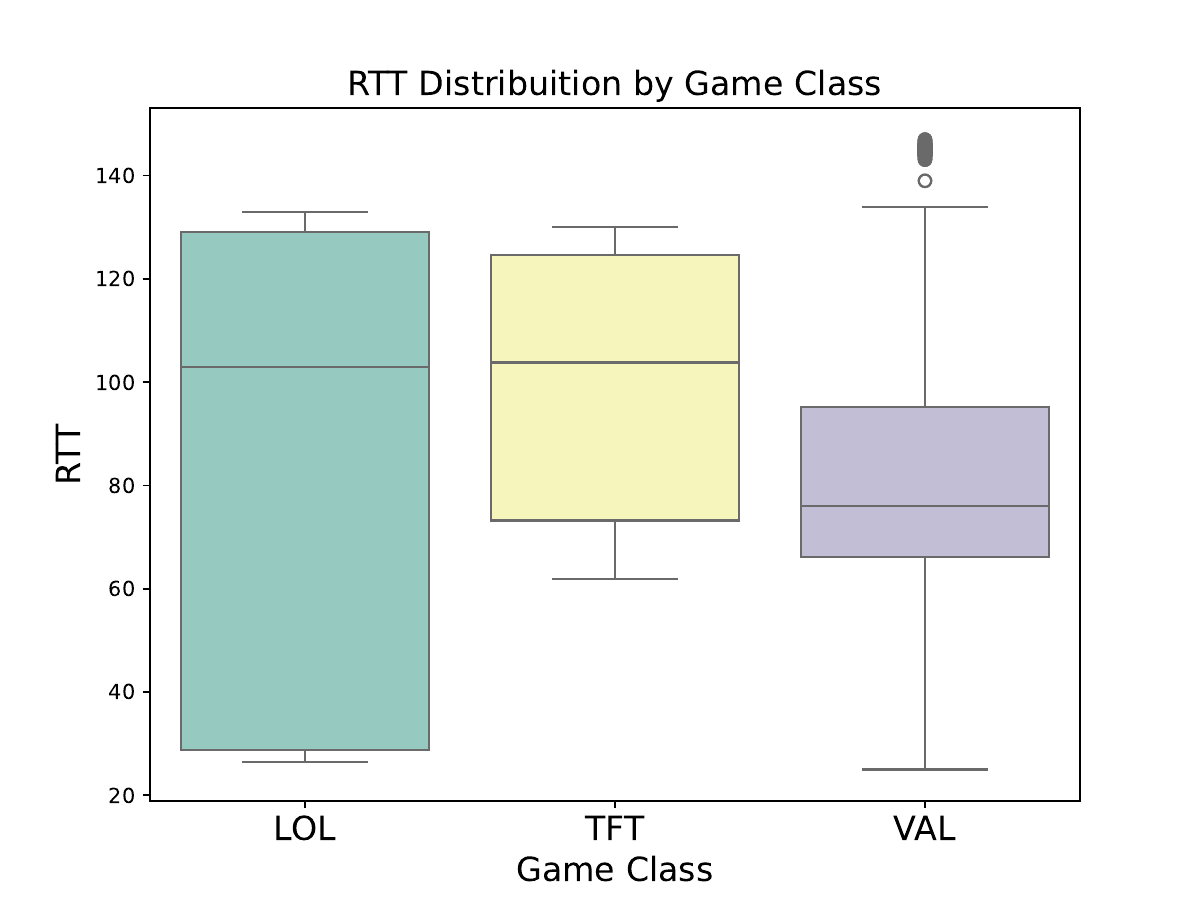} \\
        (a) How the Games are Distributed by Class. & (b) RTT Distribution. \\
    \end{tabular}
    \caption{Comparison of game distribution and RTT distribution.}
    \label{fig:combined_games_and_rtt}
\end{figure}

We analyzed the dataset's quality to train the \ac{ML} model. As shown in Figure~\ref{fig:combined_games_and_rtt}a, the class distribution is unbalanced: \ac{LOL} has the largest share with nearly 40,000 occurrences, followed by \ac{VAL} with approximately 17,000, and \ac{TFT} with approximately 15,000 games. Figure~\ref{fig:combined_games_and_rtt}b illustrates the \ac{RTT} distribution for \ac{LOL}, \ac{TFT}, and \ac{VAL}, enabling the analysis of latency variability and its potential impact on player experience.


Analyzing the interquartile range (IQR), which represents central data dispersion, \ac{LOL} exhibits the highest variability, indicating significant oscillations in network response times and potential challenges for our proof-of-concept, whereas \ac{TFT} shows a smaller IQR, suggesting more stable latency. While \ac{VAL} displays the most significant variability across the whole data range, \ac{LOL} has the highest \ac{RTT} variability in the central distribution, potentially impacting gameplay predictability. Table~\ref{tab:10_iterations} summarizes the average performance metrics of the five classification algorithms over 10 runs, including Accuracy, Precision, Recall, and F1-Score.

\begin{table}[htbp]
    \caption{Average Performance Metrics.}
    \scriptsize
    \centering
    \begin{tabular}{cccccc} 
        \hline
        \textbf{Algorithm} & \textbf{Accuracy} & \textbf{Precision} & \textbf{Recall} & \textbf{F1-Score} \\ 
        \hline
        KNN      & 0.9446 & 0.9441 & 0.9446 & 0.9443 \\ 
        RF       & 0.9481 & 0.9478 & 0.0981 & 0.9479 \\ 
        DT       & 0.9430 & 0.9429 & 0.9430 & 0.9430 \\ 
        CatBoost & \cellcolor{blue!05} 0.9483 & 0.9476 & 0.9483 & 0.9477 \\ 
        LSTM     & 0.9077 & 0.9084 & 0.9077 & 0.9057 \\ 
        \hline
    \end{tabular}
    \label{tab:10_iterations}
\end{table}

\begin{figure}[htbp]
    \centering
    \includegraphics[width=0.45\textwidth]{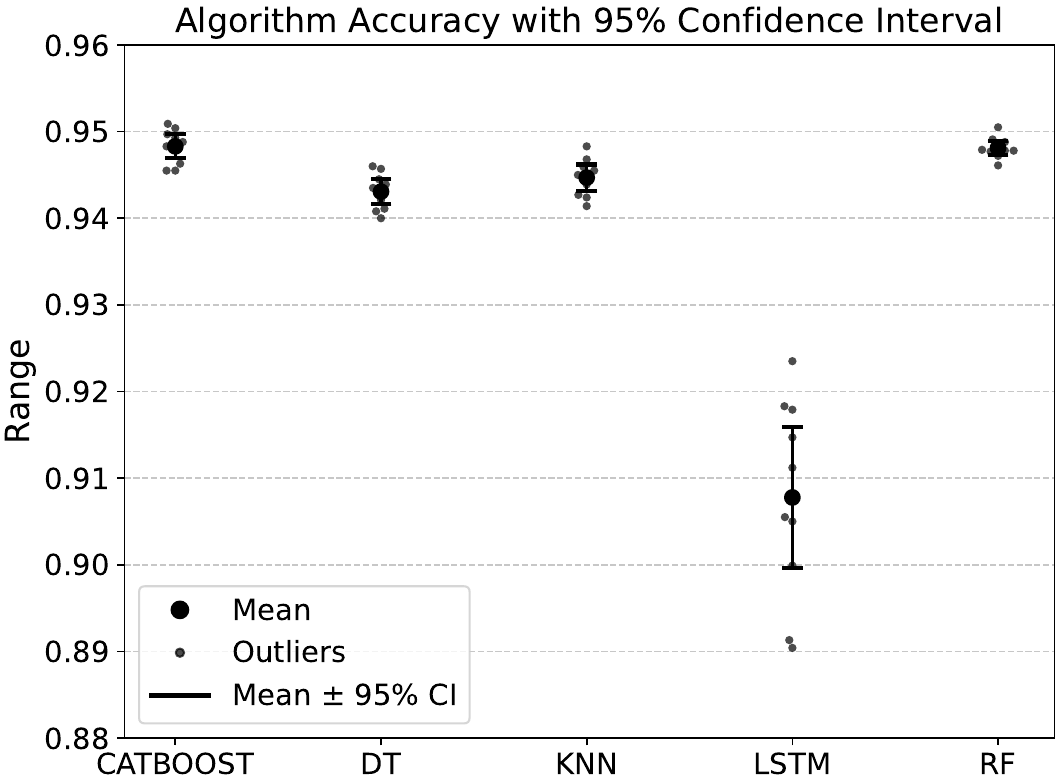}
    \caption{Accuracy Confidence Interval.}
    \label{fig:confidence_interval}
\end{figure}

The CatBoost algorithm achieved the best overall performance, with an accuracy of 0.9483 and an F1-Score of 0.9477, surpassing all other models across metrics. At the same time, Random Forest performed similarly (accuracy: 0.9481, F1-Score: 0.9479) but exhibited a low Recall (0.0981), indicating possible class imbalance. Figure~\ref{fig:confidence_interval} shows that decision tree-based models (CatBoost, \ac{DT}, \ac{RF}) and \ac{KNN} demonstrated high stability and consistent accuracy with low variability, whereas \ac{LSTM} exhibited significant variability, marked by a wide confidence interval and multiple outliers, emphasizing its sensitivity to hyperparameters and data structure; thus, decision tree-based models are more reliable for predictable applications, while \ac{LSTM} may require adjustments to improve stability.

\begin{figure}[htbp]
    \centering
    \begin{tabular}{ccc}
        \includegraphics[width=0.3\textwidth]{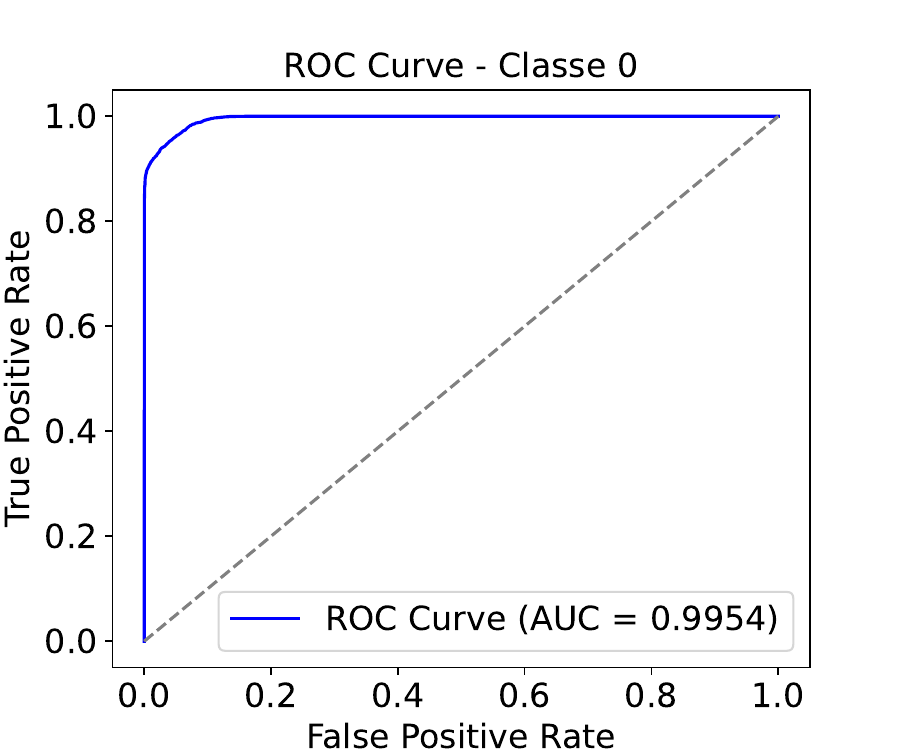} &
        \includegraphics[width=0.3\textwidth]{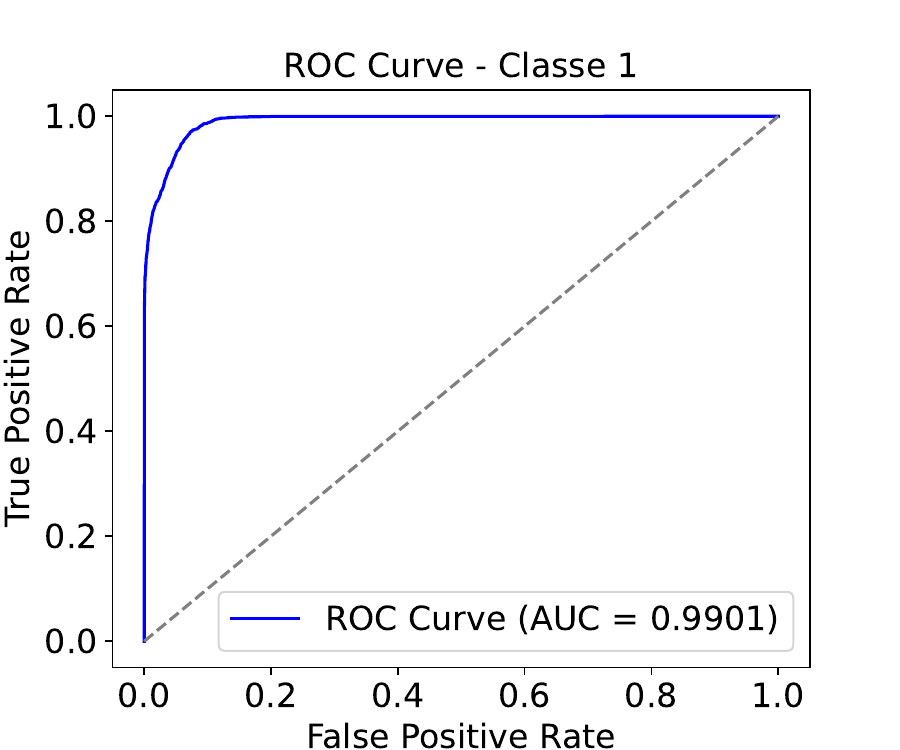} &
        \includegraphics[width=0.3\textwidth]{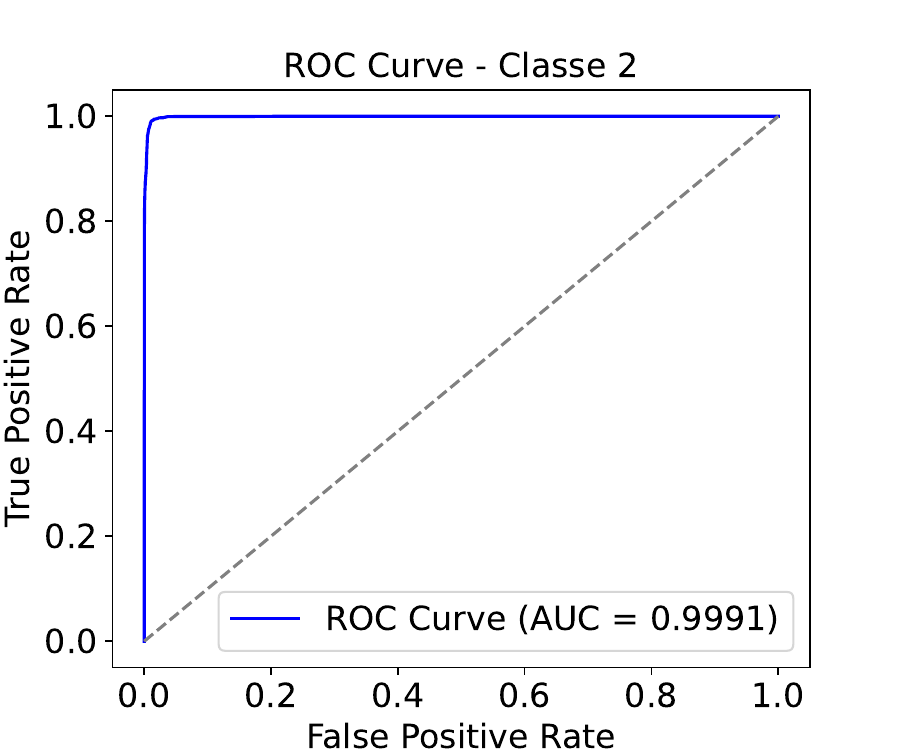} \\
        (a) LOL & (b) TFT & (c) VAL \\
    \end{tabular}
    \caption{\ac{ROC} Curve of CatBoost}
    \label{fig:roc_catboost}
\end{figure}

Figure~\ref{fig:roc_catboost} shows the \ac{ROC} curves for the three classes analyzed by the CatBoost model: (a) \ac{LOL}, (b) \ac{TFT}, and (c) \ac{VAL}. These curves evaluate the binary and multiclass classification models and represent the true positive rate (sensitivity) against the false positive rate.

\begin{figure}[htbp]
    \centering
    \begin{tabular}{ccc}
        \includegraphics[width=0.3\textwidth]{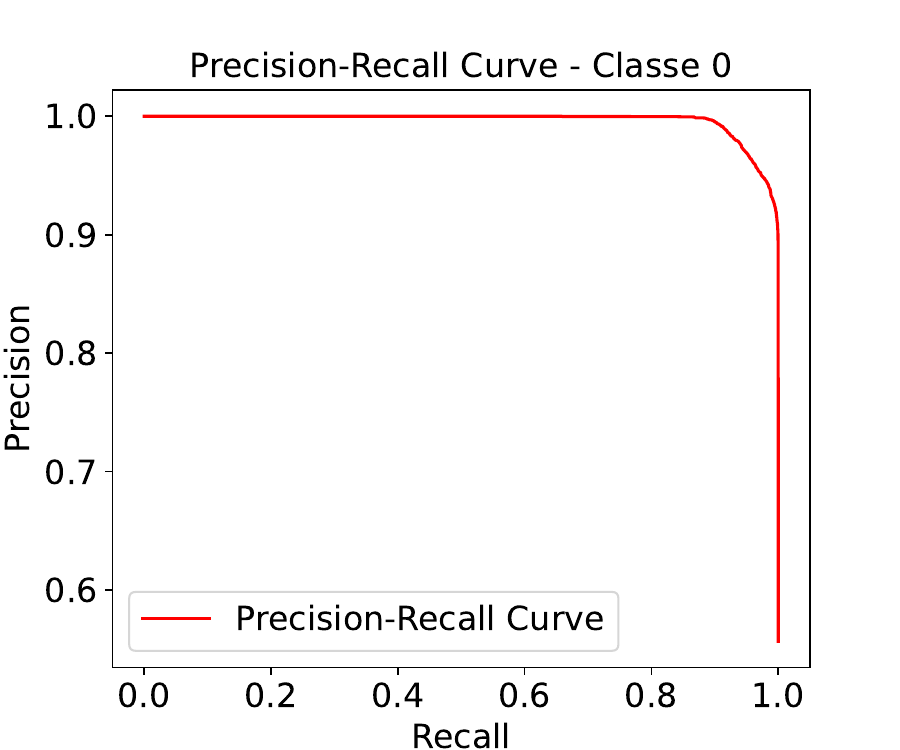} &
        \includegraphics[width=0.3\textwidth]{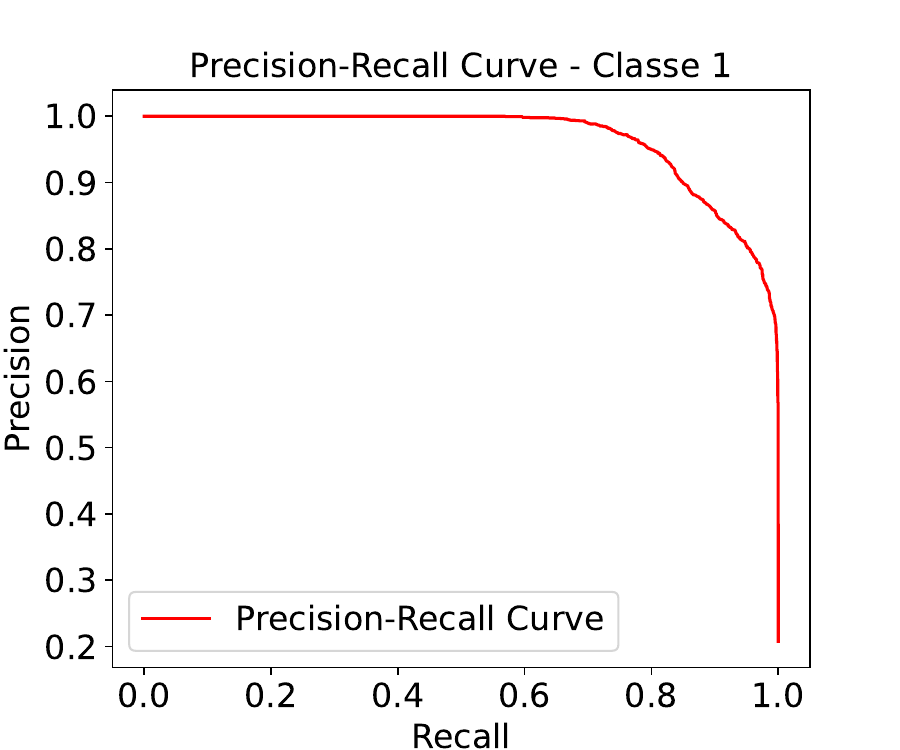} &
        \includegraphics[width=0.3\textwidth]{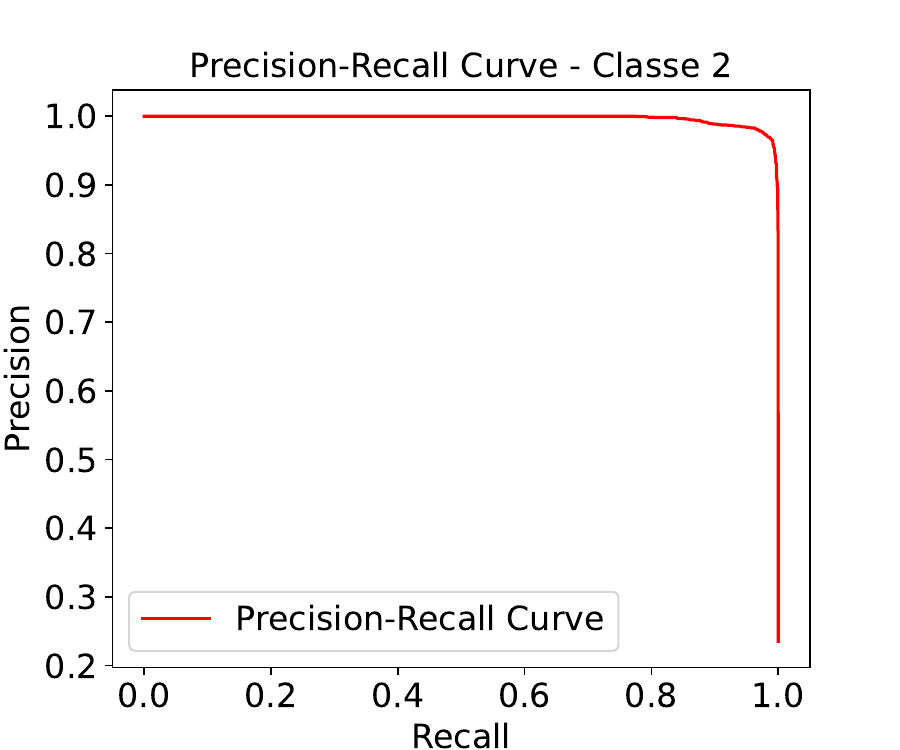} \\
        (a) LOL & (b) TFT & (c) VAL \\
    \end{tabular}
    \caption{Precision-Recall Curve of CatBoost}
    \label{fig:precisionrecall}
\end{figure}

The curves in Figure~\ref{fig:precisionrecall} exhibit strong class distinction near the upper-left corner, indicating high performance, with \ac{AUC} values exceeding 0.99 for all classes, demonstrating the model's accuracy and robustness. An \ac{AUC} near 1.0 signifies excellent discriminatory capability, while values around 0.5 indicate chance-level performance. The precision-recall curves for the three analyzed classes—(a) \ac{LOL}, (b) \ac{TFT}, and (c) \ac{VAL})—further assess the CatBoost model's performance in imbalanced scenarios, providing insight into the trade-off between precision (correct positive predictions) and recall (accurate identification of positive instances).

All curves exhibited high precision across varying recall levels, reflecting the strong classification accuracy of the model. Minor drops in precision occur only at extreme recall values, which is typical for highly reliable classification. These findings highlight the suitability of our framework for proposing user-plane interventions to enhance the \ac{UE} quality of experience in online gaming sessions.

\section{Concluding Remarks}\label{sec:concluding_remarks}

This paper presents a method to enhance edge gaming in \ac{5G} networks by instrumenting the \ac{UPF} with a user-space filter capable of estimating \ac{UE} latency using a time-shift approach. While existing approaches focus on improving cloud gaming across various network segments, our method introduces an innovative closed-loop framework between the \ac{UPF} and \ac{NWDAF}, enabling real-time latency estimation and creating new opportunities for enforcing \ac{QoS} policies.

Future work will enhance the \ac{SMF} and \ac{PCF} to dynamically adapt session parameters based on user experience, enabling real-time feedback integration and optimization for latency-sensitive applications. We also aim to evaluate performance in wired and hybrid setups, and explore AI-driven methods for predictive network adjustments and adaptive resource allocation in mobile edge environments.

\section*{Acknowledgments}

We acknowledge the financial support of the FAPESP MCTIC/CGI Research project 2018/23097-3 and FAPEMIG (Grant APQ00923-24). We also thank the FCT – Fundação para a Ciência e Tecnologia within the R\&D Unit Project Scope UID/00319/Centro ALGORITMI (ALGORITMI/UM) for partially supporting this work.

\bibliographystyle{sbc}
\bibliography{sbc-template}

\end{document}